# Kinetic Control of Recombination in Organic Photovoltaics: The Role of Spin

Akshay Rao[1], Philip C.Y. Chow[1], Simon Gélinas[1], Cody W. Schlenker[2], David S. Ginger[2], Chang-Zhi Li[3], Hin-Lap Yip[3], Alex K-Y. Jen[2,3], Richard H. Friend[1]

[1] Cavendish laboratory, University of Cambridge, Cambridge, UK
[2] Department of Chemistry, University of Washington, Seattle, Washington, USA
[3] Department of Materials Science & Engineering, University of Washington, Seattle, Washington, USA

Correspondence to: rhf10@cam.ac.uk

# Introductory paragraph

In photovoltaic diodes recombination of photogenerated electrons and holes is a major loss process. Biological light harvesting complexes (LHCs) prevent recombination via the use of cascade structures, which lead to spatial separation of charge-carriers[1]. In contrast, the nanoscale morphology and high charge densities in organic photovoltaic cells (OPVs) give a high rate of electron-hole encounters, which should result in the formation of spin triplet excitons, as in organic light emitting diodes (OLEDs)[2]. OPVs would have poor quantum efficiencies if every encounter lead to recombination, but state-of-the-art OPVs demonstrate near-unity quantum efficiency[3]. Here we show that this suppression of recombination can be engineered through the interplay between spin, energetics and delocalisation of electronic excitations in organic semiconductors. We use time-resolved spectroscopy to study a series of model, high efficiency polymer-fullerene systems in which the lowest lying molecular triplet exciton ($T_1$) (on the polymer) lies below the intermolecular charge transfer state (CT). We observe the formation of $T_1$ states following bimolecular recombination, indicating that encounters of spin-uncorrelated electrons and holes generate CT states with both spin singlet ($^1CT$) and spin triplet ($^3CT$) characters. We show that triplet exciton formation can be the major loss mechanism in OPVs. However, we find that even when energetically favoured, the relaxation of $^3CT$ to $T_1$ can be strongly suppressed, via control over wavefunction delocalisation, allowing for the dissociation of $^3CT$ back to free changes, thereby reducing recombination and enhancing device performance. Our results point towards new design rules for artificial photo-conversion systems, enabling the suppression of electron-hole recombination, and also for OLEDs, avoiding the formation of triplet states and enhancing fluorescence efficiency.

# Main text

In LHCs, photon absorption creates molecular excited states, excitons, that are funnelled to reaction centres where they are dissociated to electrons and holes, which are separated further via a series of cascading charge transfer steps[1]. In OPVs, photogenerated excitons are dissociated at a *single* donor-acceptor heterojunction formed within a de-mixed blend of the donor (D) and acceptor (A) semiconductor[4].

The most efficient OPV systems comprise nanoscale (<5 nm) domains of pure fullerene acceptor and domains of fullerene intimately mixed with a polymer donor[5,6]. These lengths scales are smaller than the Coulomb capture radius (CCR) in organic semiconductors ($kT = e^2/4\pi\varepsilon_0\varepsilon r$), which is estimated to be as large as 16 nm at room temperature due to the low dielectric constant of these materials (≈3-4)[7]. Thus, in contrast to LHCs, electrons and holes diffusing through an OPV may encounter one another before they reach the electrodes. This is similar to standard inorganic solar cells, where bimolecular electron-hole recombination (BR) determines solar cell performance.

The rate of electron-hole encounters that produce Coulombically-bound states is given by $R = \gamma np$, where $n(p)$ is the electron(hole) population density and $\gamma$ is the Langevin recombination constant given by $\gamma = q<\mu>/\varepsilon$, with $q$ the electric charge, $<\mu>$ the effective electron/hole mobility and $\varepsilon$ the dielectric constant[7,8]. This model successfully describes the principal operating mechanism of OLEDs, where charges injected through the electrodes capture one another to form strongly-bound excitons[2]. However, in empirically-optimised OPVs the recombination rate is suppressed by up to three orders of magnitude compared to the Langevin rate, allowing external quantum efficiencies (EQE) as high as 80%[3,7-9]. Here we demonstrate that the recombination of bound states formed via electron-hole encounters is mediated not only by energetics, but also by spin and delocalisation, allowing for free charges to be reformed from these bound states thus supressing recombination.

In order to probe the dynamics of these bound states, we first consider the initial dissociation of the photogenerated singlet exciton, $S_1$, at the D-A interface (Figure

1a). The first step of this process is charge transfer across the D-A interface, which can lead to either long range charge separation or the formation of bound interfacial charge transfer (CT) states[4]. Such bound charge pairs then decay to the ground state via geminate recombination (GR). It is important to note that spin must be taken into account when considering CT states as they can have either singlet ($^1$CT) or triplet ($^3$CT) spin character, which are almost degenerate in energy. Dissociation of photogenerated singlet excitons leads to the formation of only $^1$CT states due to spin conservation. In contrast, recombination of spin-uncorrelated charges should lead to the formation of $^1$CT and $^3$CT states in a 1:3 ratio based on spin statistics. $^1$CT states can either dissociate or recombine to the ground state either via luminescence (which is slow for this inter-molecular D-A process) or non-radiative decay[10,11]. For $^3$CT states, decay to the ground state is spin-forbidden and hence both radiative and non-radiative processes are very slow. However, if the energy of the lowest-lying molecular triplet exciton ($T_1$) lies below the $^3$CT energy (as is required to maximise open circuit voltage, $V_{OC}$)[12,13], then $^3$CT can relax to $T_1$. These processes are illustrated in Figure 1a.

This model for recombination and the importance of spin statistics are well established in OLEDs where the formation of (non-luminescent) triplet excitons is a major loss mechanism[2]. Efforts to overcome this problem have focused on the use of metal-organic complexes to induce spin-orbit coupling[14] and recently on the use of low exchange energy materials that can promote intersystem crossing from $T_1$ to $S_1$[15]. In contrast, for OPVs electron-hole encounters have been thought of as terminal recombination events and the roles of spin and the nature of the intermediate bound states (CT states) formed after electron-hole capture have not been explored.

Figure 1b shows the structures of the two polymers and three fullerene derivatives used as electron donors and acceptors respectively in this study. $PC_{60}BM$/ $PC_{70}BM$ (phenyl-$C_{60/70}$-butyric acid methyl ester) and ICMA ($C_{60}$ indene-fullerene-mono-adducts) are mono-substituted fullerene derivatives[16]. The lowest unoccupied molecular orbital (LUMO) of ICMA is raised less than 0.1eV in comparison to PCBM, while ICBA ($C_{60}$ indene-fullerene-bis-adducts) is a bis-substituted derivative in which the LUMO level is raised by about 0.2eV[17,18]. PCBM is the ubiquitous acceptor for OPVs while ICBA's raised LUMO is expected to boost the maximum obtainable $V_{OC}$. The donor copolymer PIDT-PhanQ (poly(indacenodithiophene-co-phenanthro[9,10-b]quinoxaline)) was chosen for this study as the spectral signatures of hole polarons and triplets are significantly different[18,19]. As we develop below, this allows us to resolve temporally the inter-conversion between polarons and triplets. It has been recently demonstrated that 1:3 blends of PIDT-PhanQ with $PC_{60}BM$ give excellent photovoltaic performance with internal quantum efficiencies (IQE) >80% and power conversion efficiencies (PCE) above 4%[18]. In contrast, blends with either ICMA or ICBA give lower performance with PCEs of 2.9%. PCPDTBT (poly[2,6-(4,4-bis-(2-ethylhexyl)-4H-cyclopenta[2,1-b;3,4-b′]-dithiophene)-alt-4,7-(2,1,3-benzothiadiazole)]) is a widely studied low band-gap polymer, which has been designed to have optimal band offsets against PCBM. However, despite extensive research the performance of PCPDTBT:$PC_{70}BM$ blends remains modest with EQE ≈ 50%[20]. Blends with ICMA and ICBA demonstrated even lower performance. Absorption spectra, EQE and J-V curves are provided in the SI.

For all the studied blends the energy of the CT state lies above $T_1$. For PIDT-PhanQ blends the energy of the CT states has been previously established using their weak photoluminescence and were found to lie at 1.31 eV, 1.36eV and 1.44eV for the PIDT-PhanQ:PCBM, PIDT-PhanQ:ICMA and PIDT-PhanQ:ICBA respectively[18]. The $T_1$ level of PIDT-PhanQ has been established to lie at 1eV[18]. The $T_1$ level of PCPDTBT has also been shown to lie at 1eV[21]. The CT state energy of PCPDTBT-PCBM blends has previously been measured to lie at 1.2eV[22]. The CT state energies of PCPDTBT-ICMA and PCPDTBT-ICBA thus lie above this value. The energy of the $T_1$ state on the fullerene derivatives lies at about 1.5eV[23]. Thus the molecular triplet exciton on the donor polymer is the lowest lying excited state for all the studied blends. We note that this is the standard configuration in the current generation of donor-acceptor systems, driven by the need to maximises $V_{OC}$, which mandates that the CT level lie close to $S_1$[12,13].

For the polymer-fullerene blends studied here steady-state photoinduced absorption (PIA) measurements have indicated the presence of long-lived triplet excitons in PIDT-PhanQ:ICBA at room temperature[18] and PCPDTBT:PCBM at low temperature[21]. However, these measurements do not reveal the mechanism for triplet formation, which has historically been considered to be via early time geminate recombination[12,24].

Here we investigate thin films of these blends using transient absorption (TA) spectroscopy. In this technique a pump pulse generates photoexcitations within the film. At some later time, we interrogate the system using a broadband probe pulse. Although TA has been widely employed to study the photophysics of OPV blends previous measurements have been severally limited by three factors. The first limitation has been insufficient temporal range, typically a maximum of 2ns delay between pump and probe. A second has been limited spectral range and lack of broadband probes, which hinders the observation of dynamic interactions between excitations. And lastly, insufficient sensitivity, which mandates the use of high fluence pump pulses to create large signals. Here we overcome these problems by using broad temporal (up to 1ms) and spectral windows (out to 1500nm) and high sensitivity (better than $5\times10^{-6}$). This temporal window is created by using an electrically delayed pump-pulse and allows for the study of long-lived charges and triplet excitons. In conjugated polymers local geometrical relaxation around charges (polaron formation) causes rearrangement of energy levels[25], bringing states into the semiconductor gap and giving rise to strong optical transitions 700nm-1500nm[26]. The absorption bands of singlet and triplet excitons are also found to lie in the near IR making a broadband spectral window necessary to track the evolution of the excited state species. The high sensitivity of the experiment is essential as it allows us to probe the dynamics of systems when the excitation densities are similar to solar illumination conditions ($10^{16}$-$10^{17}$ excitations/cm$^3$)[27]. At higher excitation densities bimolecular exciton-exciton and exciton-charge annihilation processes can dominate, creating artefacts, making such measurements unreliable indicators of device operation under AM1.5G illumination[28]. We further combine these measurements with advanced numerical techniques that allow us to resolve the spectral signatures of the overlapping excited state features and track their kinetics[29].

Figure 2a shows the TA spectra of a 1:3 PIDT-PhanQ:PCBM blend. A broad PIA feature is formed between 1100-1500nm within the instrument response (2ns). The PIA decays over several hundred ns and no spectral evolution is detected. The long lifetime of the signal and the fact that it is not observed in pristine films of PIDT-

PhanQ (see SI) rules out PIA from singlet excitons. In contrast, efficient photo generation of charge is expected in this blend and thus the PIA is assigned to the hole-polaron on the polymer. Figure 2b shows equivalent spectra for a PIDT-PhanQ:ICMA blend. Here at the earliest times the shape of the PIA is similar to that for PIDT-PhanQ:PCBM (2a), but at later times we observe significant spectral evolution. Between 1100-1170nm the signal decays with time. However, between 1300-1500nm the PIA increases for the first 50ns. The spectrum is also seen to broaden and red-shift. This spectral evolution is even more pronounced in the P1-ICBA blend shown in figure 2c. Thus unlike the PIDT-PhanQ:PCBM spectrum which shows no spectral evolution and is consistent with the decay of a single excited state, the PIDT-PhanQ:ICMA and PIDT-PhanQ:ICBA spectra suggest that a second excited state with an overlapping PIA is being formed on time scales of 10-100s of ns.

Figure 2e compares the normalised kinetics of PIDT-PhanQ:PCBM and PIDT-PhanQ:ICBA blends. The PIDT-PhanQ:PCBM blend (circles) shows no difference between the kinetics for the 1100-1200nm and 1400-1500nm regions, supporting the presence of only one excited state species. In contrast, for the PIDT-PhanQ:ICBA blend (squares) a large difference in kinetics for the two regions is observed. The rise time of the low energy region is significantly longer than for the higher energy region, indicating the growth of a second long-lived excited state species on ns time scales.

Figure 2f compares the normalised kinetics of the 1400-1500nm region in PIDT-PhanQ:ICBA as a function of pulse fluence. A clear dependence on pulse fluence is observed with rise times (to the signal maximum) as large as 80ns. Similar fluence dependence for the rise time is not observed for the 1100-1200nm region (see SI) with the signal maximum occurring within the rise time of the instrument. The rise time of the 1400-1500nm region is also fluence dependent in PIDT-PhanQ:ICMA but not for PIDT-PhanQ:PCBM (see SI). This fluence dependence in ICBA and ICMA blends indicates that the second excited state species growing in is formed via bimolecular processes.

The overlapping spectrum of the excited states makes the analysis of their kinetics difficult. In order to overcome this problem we use a genetic algorithm (GA)[29], which enables us to extract the individual spectra and kinetics from the data set. Within this approach a linear combination of two or more spectra and associated kinetics are taken and 'evolved' until they best fit the experimental data. Figure 2d shows the two spectra (solid lines) that the algorithm extracts from the PIDT-PhanQ:ICBA spectrum shown in figure 2c. The spectrum in blue is the hole-polaron and the one in red is the triplet exciton on PIDT-PhanQ. These assignments are based on previous cw PIA experiments that have established the spectral signatures of both polarons and triplets[18] as well as early time measurements of the blends following photogeneration of charges (see SI).

From the spectra and kinetics presented in figure 2 we can now observe that polarons are formed within the instrument response (2ns) in all blends. For the 1:3 PIDT-PhanQ:PCBM blend presented in figure 2a polarons then decay on a 1μs time scale and no triplet formation is observed. For PIDT-PhanQ:ICMA and PIDT-PhanQ:ICBA triplet excitons are formed via bimolecular recombination on ns time scales before decaying.

Figure 3a-b shows the kinetics, at various fluences, extracted from the GA based global analysis for PIDT-PhanQ:ICMA and PIDT-PhanQ:ICBA blends. Normalised fluence dependence of the polaron and triplet kinetics can be found in the SI. The extracted kinetics clearly demonstrate that triplets grow in as charges decay. We consider that the primary decay channel for triplets is triplet-charge annihilation, due to the high charge densities present, and model the time evolution of the system with the equation given below

$$\frac{dN_T}{dt} = -\alpha \frac{dp}{dt} - \beta [N_T][p]$$

where:

$p$: charge concentration; $N_T$: triplet concentration; $\alpha$: is the fraction of decaying charges that form triplets; β: is the rate constant for triplet-charge annihilation.

The model, solid lines in figure 3a-b, is in good agreement with the experimental data and supports the assignment of triplet formation mediated by $^3$CT following bimolecular recombination of charges. Values for the various parameters obtained from the model are summarised in the SI along with a more detailed discussion of the modelling. Values of β vary by a factor of 2 with fluence, and at values comparable to solar illumination conditions we obtain a value of 0.58 for $\alpha$ for β of $2.2\times10^{-10}$ cm$^3$/s (see SI). This demonstrates that a large fraction of the decaying charges form triplet excitons. Once formed triplets are quickly quenched due to triplet-charge annihilation. This is important: given sufficient time triplets could be re-ionised via thermal excitation to CT states. However the presence of a strong triplet-charge annihilation channel means that recombination to triplets is a terminal loss and makes it a major loss pathway in OPVs.

We now turn to the question of whether the time taken for relaxation from $^3$CT to $T_1$, process 4 shown in figure 1a with an associated timescale $\tau_4$, is fast and if not, whether there are the competing processes for the decay of $^3$CT. As noted earlier for all PIDT-PhanQ:fullerene blends the CT energy lies above $T_1$, making relaxation from $^3$CT to $T_1$ energetically favoured. However, for the more efficient 1:3 PIDT-PhanQ:PCBM blend no triplet formation is observed at room temperature (Figure 2a). But at low temperatures (<240K) bimolecular triplet formation is observed in this blend as shown in figure 3c (temperature dependent kinetics of the raw data are provided in the SI). The solid lines are fits using the model described in equation 1. This result suggests that there is a thermally activated process that competes with relaxation to $T_1$. We consider this process to be the dissociation of $^3$CT back to free changes. This is based on the fact that no other excited state species are observed for this system (Figure 2). Thus at high temperatures (>240K) the dissociation of $^3$CT back to free charges, process 3 shown in figure 1a with an associated timescale $\tau_3$, out competes relaxation of $^3$CT to $T_1$ i.e. $\tau_4>\tau_3$. Hence one of the two channels for recombination (the other being recombination through $^1$CT) is supressed, allowing for high EQEs. At lower temperatures this dissociation process is supressed, such that $\tau_4<\tau_3$, leading to a build up of triplet excitons as seen in figure 3c.

The above analysis raises the question of why triplet formation is observed in ICBA and ICMA blends at room temperature but is out-competed by dissociation back to free charges in the 1:3 PIDT-PhanQ:PCBM blend. We noted above the CT level of ICMA and PCBM blends are within 50 meV and hence a simple energetic consideration is unlikely to explain this difference. Our previous work on CT states formed at early times via ionisation of excitons at heterojunctions suggested that their dissociation was mediated by charge delocalisation, with greater wave function delocalisation allowing for more efficient CT state dissociation[4]. We consider that the same mechanisms must be applicable to CT states formed via bimolecular recombination. It is known that PCBM crystallises efficiently and that this aids charge separation[30], most likely due to delocalisation of the electron wavefunction over the PCBM crystallites. In contrast, the substitutions on the fullerene in ICMA and ICBA prevent effective crystallisation, which would lead to more localised electron wavefunctions. This would imply that CT states formed via recombination were more loosely bound in PIDT-PhanQ:PCBM in comparison to PIDT-PhanQ:ICMA and PIDT-PhanQ:ICBA and thus more susceptible to dissociation back to free changes. To test this hypothesis we study the recombination dynamics in a 1:1 PIDT-PhanQ:PCBM blend spun from chloroform. The lower fullerene concentration and low boiling point solvent lead to a more intimate blend and arrests the growth of fullerene aggregates. Bimolecular triplet formation is observed in this blend, figure 3d, which shows the normalised fluence dependence of the triplets (raw data and polaron dynamics are shown in the SI). Thus by disrupting fullerene aggregation and hence charge delocalisation we make $\tau_4 < \tau_3$. This result confirms that delocalisation plays a crucial role in recombination.

In order to generalise these results we now turn to a study of the well-studied donor polymer PCPDTBT. Figure 4a shows the evolution of the TA spectrum of a 1:2 PCPDTBT:ICBA blend. A broad PIA feature between 1175-1550 nm is formed within the instrument response (2ns). A shifting of the peak from 1300nm to 1275nm is observed over tens of ns, before the PIA decays over several hundred ns. A blue shift was also observed in the TA spectra of PCPDTBT:PC$_{70}$BM and PCPDTBT:ICMA films (see SI).

The solid red line in figure 4b shows the triplet spectrum extracted via a GA analysis of the blends, using the measured polaron spectrum, blue line, as a reference for the algorithm. The extracted spectrum shows excellent agreement with the triplet spectrum measured via doping a neat PCPDTBT film with an iridium complex that enhances the intersystem crossing rate leading to a high triplet population, dashed red line. The triplet peak at higher energy, with respect to the polaron, explains the blue shifting of the TA spectrum in figure 4a as triplets grow in.

Figure 4c shows the fluence dependence of the polaron and triplet for a PCPDTBT:$PC_{70}BM$ film, similar to those shown in figure 3a-b. The solid lines are fits to the experimental data obtained using equation 1, and support the general applicability of the presented model. The result also explains why the PCPDTBT:$PC_{70}BM$ system displays only modest EQEs[20], with recombination to triplets being a major loss mechanism even for the $PC_{70}BM$ blend.

Based on the above analysis we can now put in place a new photophysical picture of recombination in OPVs, summarised in figure 1a. Bimolecular electron-hole capture events lead to the formation of CT states with both spin singlet and spin triplet character, $^1CT$ and $^3CT$. Recombination from $^3CT$ back to the ground state is spin forbidden, but in most systems $T_1$ lies below CT so that energetic relaxation to bound triplet excitons is favourable (with an associated time $\tau_4$). Under standard operating conditions, triplet formation is a terminal recombination channel, because the large charge population in the device rapidly quenches these triplets, preventing thermally-activated re-ionisation. However, as we show here, it is possible to use wavefunction delocalisation such that the reionisation of CT states back to free charges (with an associated time $\tau_3$) becomes competitive with relaxation to $T_1$ ($\tau_4 > \tau_3$). Delocalisation brings the system into a regime where kinetic control overrides energetic limitations, suppressing the dominant recombination channel and enhancing photovoltaic performance. Our result puts in place important design rules not only for artificial photo-conversion systems such as OPVs, but also for OLEDs where careful design of the heterojunction would suppress recombination to low lying triplets and boost fluorescence efficiency.

## Methods Summary

PCPDTBT, ICMA and ICBA were obtained from One-materials, $PC_{60}BM$ and $PC_{70}BM$ from Nano-C. PIDT-PhanQ was synthesised as described previously[18,19].

For the PIDT-PhanQ:fullerene thin film samples, 1:3 polymer:fullerene blends (20mg/ml in dichlorobenzene) were spun on fused silica substrates. The PIDT-PhanQ:PCBM thin film discussed in figure 3d was spun from a 1:1 blend (20mg/ml in chloroform).
For the PCPDTBT:fullerene thin film samples, 1:2 polymer:fullerene blends (30mg/ml in chlorobenzene) were spun on fused silica substrates.

For transient absorption (TA) measurements 90 fs pulses generated in a Ti:Sapphire amplifier system (Spectra-Physics Solstice) operating at 1 KHz were used. The broad band probe beam was generated in a home built parametric amplifier (NOPA). Pump pulses were generated by frequency-doubled *q*-switched $Nd:YVO_4$ laser (532nm). Delay times from 1 ns to 100 μs were achieved by synchronizing the pump laser with the probe pulse by use of an electronic delay generator. Samples were measured in a dynamic vacuum better than $1\times10^{-5}$ mbar.

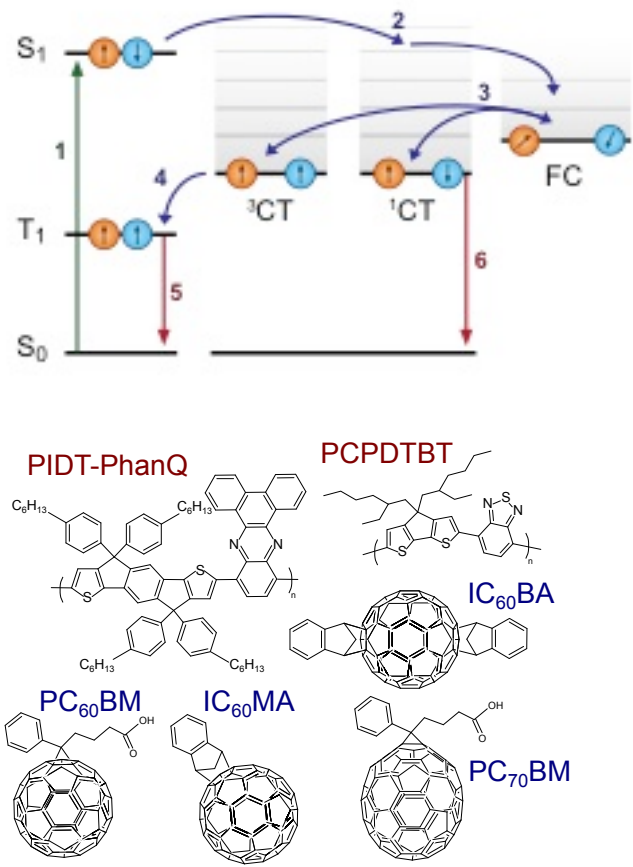

**Figure 1: (a)** State diagam representing the various photophyical process in an OPV, conversions between excited state species are shown in blue while recombintion channels are shown in red. $S_1$ and $T_1$ are the lowest lying singlet and triplet excitons respectively and CT is the chrarge transfer state. **1**: Photoexcitation creates a singlet exciton. **2**: The singlet exciton is ionised at a heterojunction leading to the formation of $^1$CT states which separate into free charges (FC) with high efficiency. **3**: Bimolecular recombination of electrons and holes leads to the formation of $^1$CT and $^3$CT states in a 1:3 ratio, as mandated by spin statistics. The $^1$CT state can recombine slowly to the ground state as shown in **6**. For the $^3$CT state recombination to the ground state is spin forbidden, but relaxation to the $T_1$ state, **4**, is energetically favourable. Once formed triplet excitons can return to the ground state via an efficent triplet-charge annhilation channel, **5**. Under favourable conditons, as developed in the text, the time required for CT states to reionise to free charges (process 3:$\tau_3$) is less than the time required for relaxation to $T_1$ (process 4:$\tau_4$). Thus CT states are recycled back to free charges leading to a supression of recombination. **(b)** Molecular structures of the donors and acceptors used in this study.

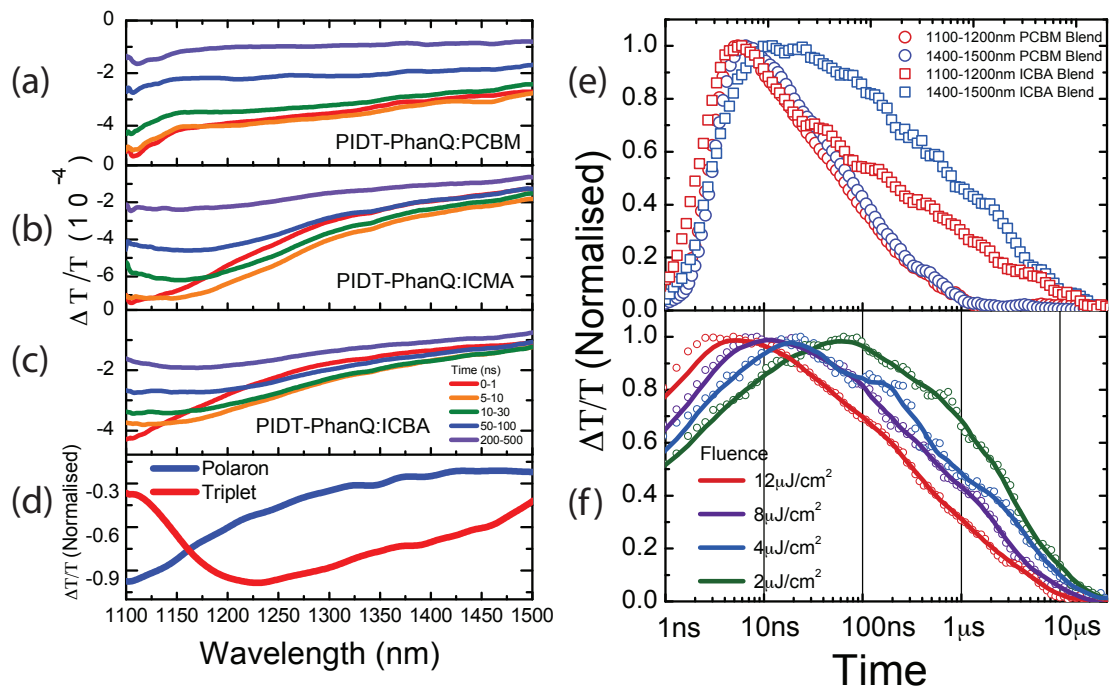

**Figure 2: (a-c)** Temporal evolution of the TA spectrum for 1:3 blend ratios of PIDT-PhanQ:PCBM, PIDT-PhanQ:ICMA and PIDT-PhanQ:ICBA respectively. Samples were excited with an excitation fluence of 8μJ/cm². Temporal slices are averaged over indicated time periods and smoothed. The PCBM blend (a) shows only a slow decay with no spectral evolution. In contrast, ICMA (b) and ICBA (c) blends show significant evolution up to 100ns, indicating the growth of a new excited state species. **(d)** The spectra extracted from the genetic algorithm (GA) analysis of the ICBA blend (c) showing the triplet and charge polaron spectra. **(e)** Comparison of kinetics of PIDT-PhanQ:PCBM (circles) and PIDT-PhanQ:ICBA (squares). Both the high energy (red) and low energy regions (blue) for PIDT-PhanQ:PCBM decay on the same time scale indicating the presence of only one excited state species. In contrast, for PIDT-PhanQ:ICBA the regions show divergent kinetics indicating the presence of multiple excited state species. **(f)** Fluence dependence of the low energy region (1400-1500nm) for PIDT-PhanQ:ICBA. The fluence-dependent growth of the feature demonstrates that the second excited state species, triplets, are formed via bimolecular processes.

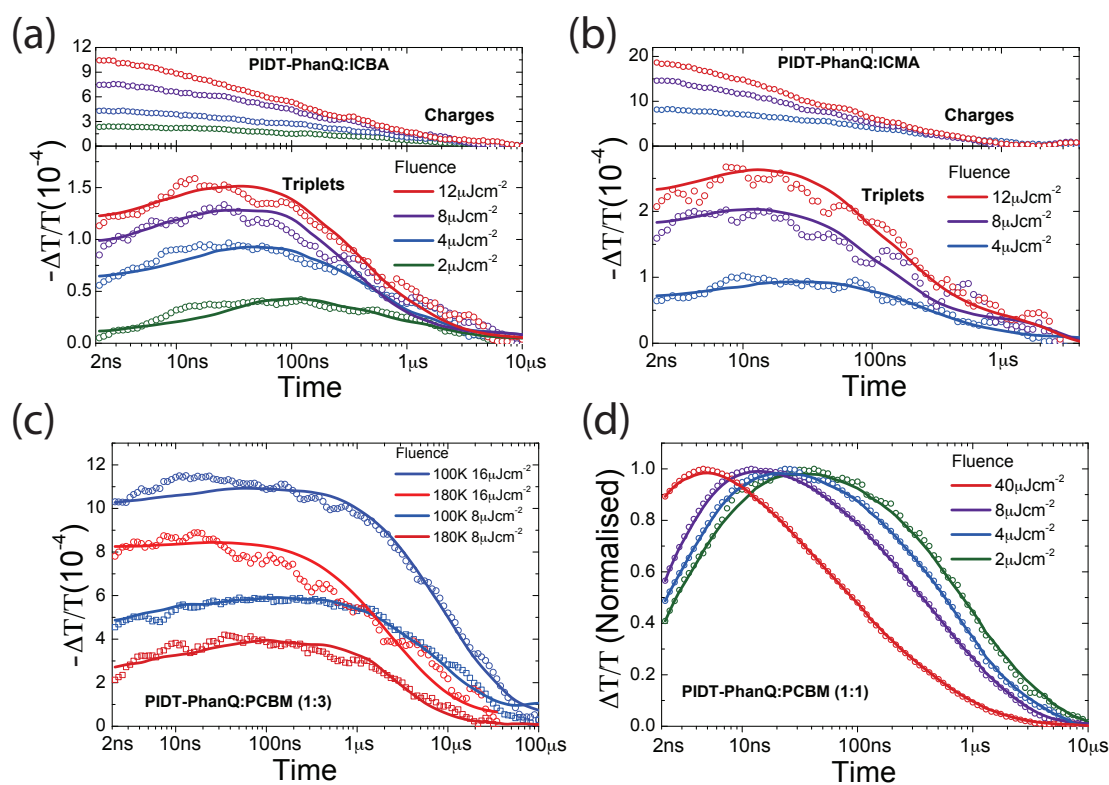

**Figure 3: (a-b)** Charge and triplet dynamics (circles) for PIDT-PhanQ:ICBA and PIDT-PhanQ:ICMA respectively extracted from the genetic algorithm (GA) analysis. Charges are formed within the instrument response in all cases. The growth of triplets is fluence dependent, with a maximum population attained at longer times for lower fluences. The solid lines are fits of the experimental data using the model described in the text. **(c)** Temperature dependent triplet dynamics (extracted by the GA analysis) for a 1:3 PIDT-PhanQ:PCBM sample. Open symbols are the experimental data and solid lines are fits using the model described in the text. The maximum triplet population is formed at longer times at lower fluences and lower temperatures consistent with a bimolecular diffusion dependent process. (d) Room temperature fluence-dependent triplet dynamics (extracted by the GA analysis) for a 1:1 PIDT-PhanQ:PCBM blend spun from chloroform. In contrast to the 1:3 PIDT-PhanQ:PCBM blend (figure 2a) this blend shows room temperature triplet formation which, as indicated by the fluence dependence, stems from bimolecular processes. Solid lines are guides to the eye.

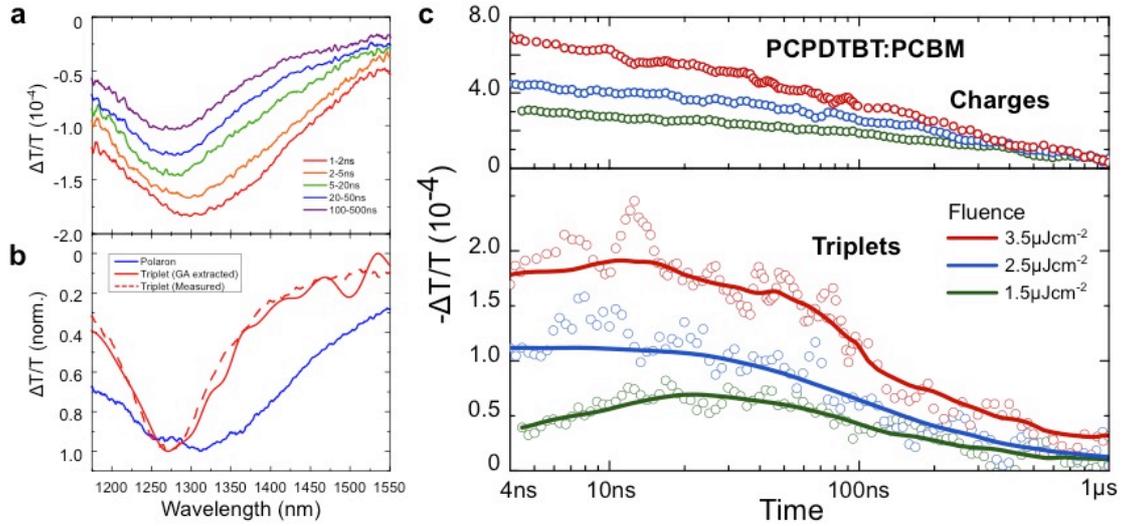

**Figure 4: (a)** Temporal evolution of the TA spectrum for PCPDTBT:PCBM, excited with an excitation fluence of $2\mu J/cm^2$. Temporal slices are averaged over indicated time periods and smoothed. A blue shift of the spectra from a peak at 1300nm to 1275nm can be seen over the first 100ns. **(b)** Triplet spectrum extracted from the GA analysis, solid red line, and that measured via doping a PCPDTBT thin film with a triplet sensitizer, dashed red line. The blue line shows the charge spectrum as measured 50ps after photoexciation, sufficient time for charge generation but before triplet formation would begin. **(c)** Charge and triplet dynamics (circles) for PCPDTBT:PCBM extracted from the global GA analysis, analogous to those shown in figure 3a-b. Charges are formed within the instrument response in all cases. The growth of triplets is fluence dependent, with a maximum population attained at longer times for lower fluences. The solid lines are fits of the experimental data using the model described in the text.